# MLPro: A System for Hosting Crowdsourced Machine Learning Challenges for Open-Ended Research Problems

Peter Y. Washington, Stanford University and MLPro

Aayush Nandkeolyar, MLPro

Sam Yang, MLPro

The task of developing a machine learning (ML) model for a particular problem is inherently open-ended, and there is an unbounded set of possible solutions. Steps of the ML development pipeline, such as feature engineering, loss function specification, data imputation, and dimensionality reduction, require the engineer to consider an extensive and often infinite array of possibilities. Successfully identifying high-performing solutions for an unfamiliar dataset or problem requires a mix of mathematical prowess and creativity applied towards inventing and repurposing novel ML methods. Here, we explore the feasibility of hosting crowdsourced ML challenges to facilitate a breadth-first exploration of open-ended research problems, thereby expanding the search space of problem solutions beyond what a typical ML team could viably investigate. We develop MLPro, a system which combines the notion of open-ended ML coding problems with the concept of an automatic online code judging platform. To conduct a pilot evaluation of this paradigm, we crowdsource several open-ended ML challenges to ML and data science practitioners. We describe results from two separate challenges. We find that for sufficiently unconstrained and complex problems, many experts submit similar solutions, but some experts provide unique solutions which outperform the "typical" solution class. We suggest that automated expert crowdsourcing systems such as MLPro have the potential to accelerate ML engineering creativity.

CCS CONCEPTS •Human-centered computing~Collaborative and social computing~Collaborative and social computing theory, concepts and paradigms~Computer supported cooperative work•Human-centered computing~Collaborative and social computing~Collaborative and social computing systems and tools•Computing methodologies~ML

## 1 INTRODUCTION

Several innovations in machine learning (ML) require creativity over mathematical prowess. This is especially true for deep learning, where the learned model is largely a black box, and the application of state-of-the-art methods is often guided by intuition rather than formal proofs or logic. For example, innovative methods for the new field of self-supervised learning often require creativity when defining a pretext task, creating data augmentations for contrastive learning, and engineering loss functions, network architectures, and synthetic data generation strategies [13, 29]. We explore the potential for crowdsourcing to facilitate a "breadth first" exploration of open-ended ML problems.

There are myriad existing crowdsourcing and citizen science platforms which distribute labor-intensive data labeling and creativity-driven scientific pursuits to both expert and non-expert populations [8]. This practice started with the SETI@Home and Folding@Home projects, where radio signals from space [2] and protein folding simulations [4] (respectively) were analyzed using the computing power of distributed computers in homes and offices around the world [2]. Several citizen science projects followed which can be grouped into problem-oriented games and data-oriented games [8]. Problem-oriented games consist of finding solutions to computationally and scientifically intractable problems using the power of the crowd. Prominent examples include molecular structure modeling on FoldIt [9], molecular design on Eterna [21], and sequence alignment on Phylo [19]. Data-oriented games involve acquiring and labeling data for science and ML. A prominent example is GalaxyZoo [22], where people label images of galaxies. GalaxyZoo was the first of a variety of data labeling games posted as part of the Zooniverse [27].

We explore the intersection of citizen science for ML with classical online code judging systems and ML integrated development environments (IDEs). The most popular ML IDEs follow a notebook development paradigm [12, 14, 26], such as Jupyter and Google Colab notebooks. While these platforms provide full flexibility for developing open-ended solutions, it is difficult to auto-grade separate notebooks, as (1) they each run as separate and independent files and (2) the coder can change any part of the notebook and write their answers in any desired format. Constraints set by the problem creator can be difficult to enforce at scale. We are also inspired by existing online code judging websites such as LeetCode, HackerRank, and Code Chef [34]. Existing online code evaluation systems are optimized for software engineering questions such as algorithm implementation and system design. A problem is formulated, and several test cases are provided to the user. To prevent users from cheating, some test cases are usually hidden from the user. When the user submits their code solution, their function or class is evaluated against all test cases. A submission is considered to "pass" only if all test cases pass. Mogavi et al. identify five user archetypes on online coding platforms platforms: "challenge-seekers", "subject-seekers", "interest-seekers", "joy-seekers", and "non-seekers" [10]. Similar coding environments have been created for domain-specific purposes such as educational settings to enable automatic coding feedback [23, 36], music [1], educational games [35], like life science education and experimentation [15, 17-18, 30-33], embedded hardware programming [25], and programming education [3, 7, 16]. We propose an ideal "middle ground" between unconstrained environments like Jupyter notebooks and highly constrained environments like test case-based grading, and we optimize for the domain of ML.

We present MLPro, a structured web environment for hosting ML problems. MLPro consists of open-ended ML coding challenges where participants are evaluated based on the performance of their method on a held-out test set rather than checking the code against test cases. MLPro allows users to submit a variety of possible solutions and approaches. Submissions are not graded on a pass/fail system, but rather based on performance on ML metrics on a held-out and hidden test dataset. We run a feasibility study of whether such a system can garner a diversity of solutions for open-ended ML problems which are crowdsourced to a large group of ML experts. We find that for sufficiently unconstrained and complex problems, many experts submit similar solutions, but some experts provide unique solutions which outperform the "average" solution. Systems like MLPro are a promising avenue for accelerating engineering experimentation and innovation for certain problems with ambiguous solutions.

**MLPRO SYSTEM**

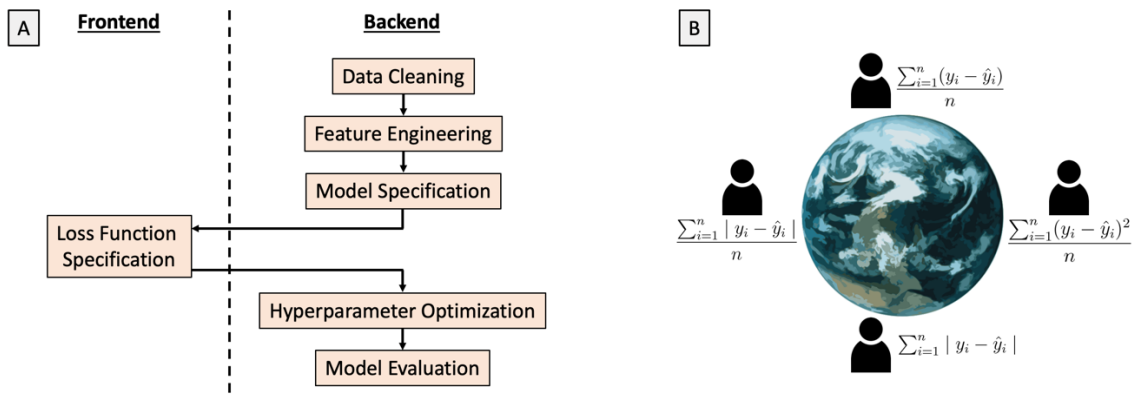

Figure 1: MLPro extracts one or more tasks in the ML development pipeline and presents it to a crowd of domain experts (ML engineers). Here, we show the example of a loss function specification challenge (A). Each distributed crowdsourced engineering worker provides a code solution (in the example presented here, different loss function implementations) for a task (B) and submits it to the backend for integration into the rest of the ML pipeline, which is hidden from the user. This paradigm allows for all users to be compared and automatically graded on their individual ML contribution.

MLPro is a web-based IDE which supports open-ended ML coding challenges to be completed by several members of a distributed crowd of engineers. Each user completes a pre-specified component of the ML development cycle, their code is plugged into a full ML pipeline hidden from the user, and the resulting performance is displayed to the user and can be used to rank independent submissions of the coding challenge. The user interface consists of 3 panels (Figure 2). The left panel describes the open-ended question which usually comprises one or two distinct steps of the ML development pipeline. In the example depicted in Figure 2, a description of a data imputation challenge for tabular data is provided. The middle panel provides a Python code editor with function stubs that must be completed by the user. For the data imputation challenge depicted in Figure 2, the user must complete a `modify_X_train()` function which takes as input the unimputed training data (`X_train`) and returns a modified version of the training data with all missing values of the data filled in. The right panel displays the user's score on ML metrics as well as any console output. In Figure 2, the accuracy, precision, and recall scores are shown when an undisclosed classifier is trained on the imputed training data provided by the user and evaluated on testing data which is hidden from the user.

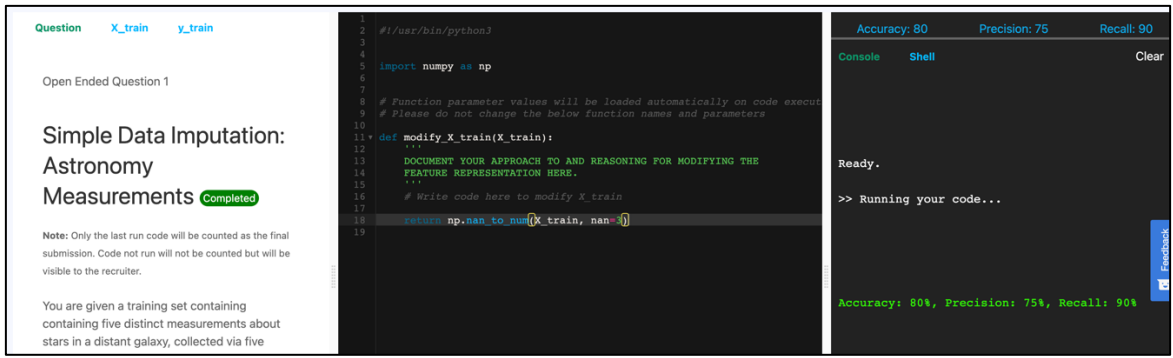

Figure 2: User interface of the open-ended challenges on MLPro. The coding challenge description is displayed on the left panel. The user's code is written in the middle panel. The user's output and corresponding score are displayed on the right panel.

MLPro can support several challenge types corresponding to a single step in the model development pipeline. We developed open-ended challenges for classification model development, regression model development, feature selection, dimensionality reduction, data imputation, loss function specification, and feature engineering. In principle, the system can support several other challenge types related to any aspect of the ML development cycle. For any challenge type, a function stub corresponding to a particular step in the ML development cycle is provided to the user. All other steps of the ML pipeline are not exposed to the user and are instead run on backend infrastructure. In a model development challenge, for example, a `create_model()` function stub is provided to the user, and the user is expected to return a model object with an implemented `fit()` function for training the model and `predict()` function for making predictions with the resulting model. In other instantiations of model development challenges, the user can be constrained to return a TensorFlow, PyTorch, or Scikit-Learn model. In feature selection, dimensionality reduction, data imputation, and feature representation challenges, the user is provided a `modify_X()` function (or similar variation). The user's functions are plugged into the full ML development pipeline in the backend, and the resulting scores on the ML metrics for that problem are sent to the client.

In each coding challenge, the user is provided with a training dataset, `X_train`, and the corresponding training labels, `y_train` (Figure 3A). A separate testing dataset (`X_test` and `y_test`) are withheld from the user on the backend (Figure 3B). After the user writes their code for the coding challenge (Figure 3C), the code is sent to the backend, where each client's code is evaluated, potentially in parallel with other submissions, on the held-out test dataset which is not visible to the challenge participant (Figure 3D). The model is then evaluated on a pre-defined set of ML metrics and the results are sent back to the user for feedback (Figure 3E).

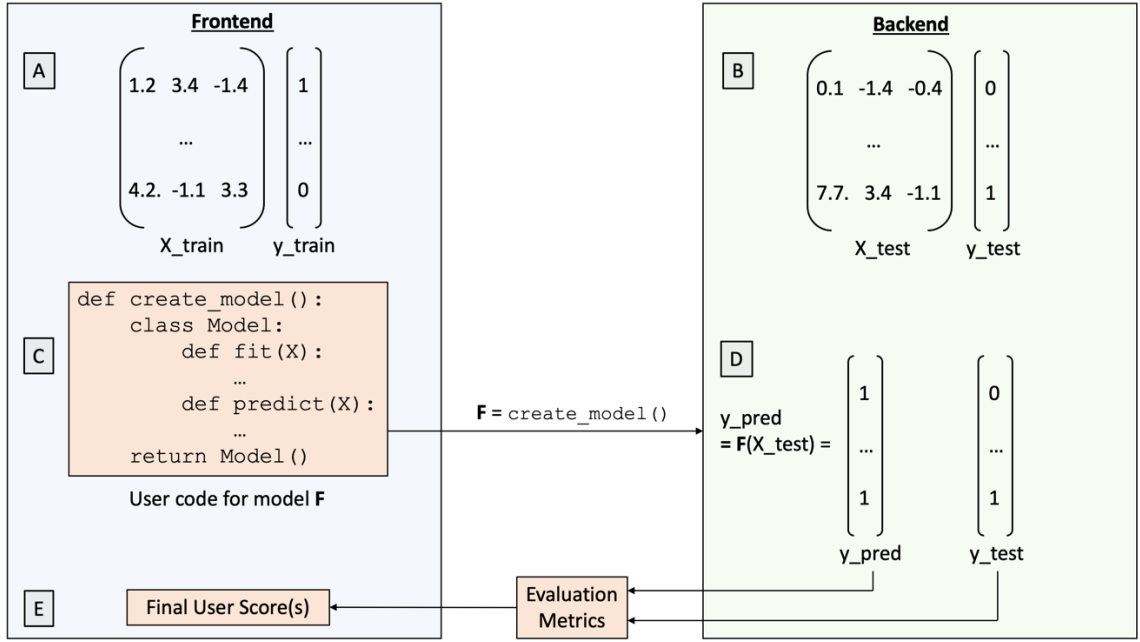

Figure 3: Process for evaluating open-ended solutions. Here, we show an example of the process for model development challenges, although MLPro supports several other challenge types. The user is provided with training data, X_train, and corresponding training labels y_train (A). A separate testing dataset, consisting of X_test and y_test, are withheld from the user (B) and used to evaluate the strength of the user's contributed method. The user's code solution (C) is sent to the backend server and used to make predictions on the held-out testing dataset (D). ML evaluation metrics appropriate for the challenge problem are calculated from the user's model predictions y_pred and the true values y_test and displayed to the user (E).

The primary system-level innovations of MLPro over traditional development environments like Jupyter and Colab notebooks are (1) the splitting of the ML development process between multiple clients and a central backend and (2) the paradigm of isolating "bite-sized chunks" of the ML lifecycle which are abstracted and crowdsourced to several engineers. This system design enables creativity and flexibility by each contributor of a crowdsourced solution while providing enough structure for automated ranking of solutions. Each client corresponds to a unique user who contributes a code submission to the ML challenge.

The open-ended nature of problems on MLPro naturally prevents cheating. While test case-centered software engineering coding platforms must rely on automated coding style and plagiarism analysis, MLPro ensures a further level of anti-cheating by the possibility of an unbounded number of solutions and approaches.

MLPro is implemented as a web application using the Django Python framework hosted on an Elastic Cloud Compute (EC2) instance on Amazon Web Services (AWS). The backend of MLPro is hosted on AWS Lambda as a serverless compute service running FastAPI. Lambda partitions a separate 10GB RAM compute service for each independent client submission, with a limit of 10,000 concurrent submissions.

## 2   USER STUDY 1: REGRESSION CHALLENGE

As an initial user study, we posted a straightforward regression problem using the popular UCI ML housing dataset [5, 11]. Challenge participants were tasked with building any type of regression model. The challenge text was as follows (verbatim):

*Your task is to build any type of ML model which can predict housing prices given some information about the house. We will use the UCI ML housing dataset. The number you are to predict is the price, in thousands of dollars, as of 1978.*

*The housing price training data variables are called X_train and y_train and can be viewed on the tabs to the left (or printed to the console). Your task is to return an appropriate regression model for these data. We will use your returned model to train on these data and evaluate on a held out test set hidden from you. You will be evaluated on the mean squared error of your model on the test set. Since this is a regression problem, lower scores are better.*

*You may either return a scikit-learn or tensorflow model, or you can create your own model class which conforms to the following API:*

```
class Model:
    def __init__(self):
        pass
    def fit(X_train, y_train):
        # Train the model.
        pass
    def predict(X_test):
        # Return predictions of the model on X_test.
        pass
```

*You do not need to train your model (that is, call the fit method of your model). We will do that for you. We have provided a baseline working solution which you should try to beat.*

*The goal of these open-ended questions is not to simply hyperparameter tune and debug the existing solution. We provide an existing solution as a not-so-great starting place. Using this as an example, the goal should be to come up with a much better solution from scratch.*

To qualify for the challenge, potential participants were required to pass a series of multiple choice and testcase-based coding questions related to fundamental machine learning concepts. Only participants who scored a 100% were eligible to participate in this study. 20 potential participants qualified for inclusion in the user study.

Table 1 displays the approaches developed by users in the data imputation challenge. Similar approaches are grouped into the same category, and most solutions are not unique. The majority of participants either used linear regression or a decision tree-based regression approach to solve the problem. One participant used support vector regression. All participants elected to use the Scikit-learn [24] Python library rather than implementing their own solution. Given the lack of unique solutions in our initial study, we do not elaborate on the provided solutions here.

| Approach | Count | Mean MSE (± Std.) |
| --- | --- | --- |
| Linear regression (with or without regularization) | 6 | $33.72 \pm 22.88$ |
| Decision tree-based regression | 13 | $9.44 \pm 0.80$ |
| Support vector regression | 1 | 11.13 |

Table 1: Approaches submitted by users in the regression challenge. Most solutions were either linear regression or a tree-based regression method. One participant used support vector regression. We conjecture that unique solutions were not provided by participants given the simplicity of both the problem and the dataset.

## 3  USER STUDY 2: IMAGE DIMENSIONALITY REDUCTION CHALLENGE

Given the lack of diversity in provided solutions in User Study 1, we wanted to next evaluate MLPro with a more open-ended challenge. We posted an image dimensionality reduction challenge where the goal was to minimize the dimensions of input images while maximizing the predictive power of the reduced representation size. The challenge text was as follows (verbatim):

*You have a dataset containing images of handwritten digits. Your goal is to reduce the dimensionality of each image to be a vector of length 20 or less so that the model can be fast and not prone to overfitting due to the small dataset.*

*We will train and evaluate a model using your modified dataset.*

*The handwritten digits data variables are called X_train and y_train and can be viewed on the tabs to the left (or printed to the console). You will be evaluated on the multi-class accuracy, precision, and recall of your model on the test set.*

*The X_train variable is a NumPy array containing 8 x 8 images. The y_train variable contains numerical labels. While you cannot change the label representation in this problem, you may (and must) change the input representation to be less than 20 dimensions, otherwise you will receive a score of 0.*

*Important: You will receive a ValueError if you don't modify the input images to be 1-dimensional vectors rather than 2-dimensional images.*

*The goal of these open-ended questions is not to simply hyperparameter tune and debug the existing solution. We provide an existing solution as a not-so-great starting place. Using this as an example, the goal should be to come up with a much better solution from scratch.*

To qualify for the challenge, potential participants were required to pass a series of multiple choice and testcase-based coding questions related to fundamental machine learning concepts. Only participants who scored a 100% were eligible to participate in this study. 47 potential participants qualified for inclusion in the user study.

Table 2 displays the approaches submitted by users in the image dimensionality reduction challenge. 28 users submitted a standard dimensionality reduction algorithm such as principal component analysis, t-SNE, or SVD [6, 28]. 13 users submitted a downsampling method. 6 users submitted unique non-standard solutions. Table 2 summarizes the performances.

Of the unique solutions, two solutions (feature hashing and center-cropping) resulted in test performances above one standard deviation from the mean of the standard solutions. One solution (taking the mean pixel value of image regions) resulted in performance on par with the mean of the better-performing standard solution (downsampling). Three unique solutions resulted in sub-par performance.

| Approach | Count | Mean MSE (± Std.) |
|---|---|---|
| Standard dimensionality reduction algorithm | 28 | 25.71 ± 17.36 |
| Downsampling | 13 | 80.80 ± 8.65 |
| Unique: Feature hashing | 1 | 90.28 |
| Unique: Center-crop the image | 1 | 91.62 |
| Unique: Split image into sections and take the mean of each section | 1 | 82.91 |
| Unique: Deep learning autoencoder | 1 | 9.88 |
| Unique: Linear algebra image manipulation | 1 | 48.74 |
| Unique: Local Binary Pattern descriptor, contour based | 1 | 60.47 |

Table 2: Approaches employed by users in the image dimensionality reduction challenge.

## 4 DISCUSSION AND CONCLUSION

We present a system, MLPro, which supports distributed and concurrent submission of semi-structured solutions to open-ended ML problems. When evaluating MLPro with ML practitioners, we found that some users submitted unique solutions which outperformed the standard classes of solutions provided by the majority of participants. While we ran only a feasibility study, there is potential for systems like MLPro to foster exploration and creativity in ML-related problems and possibly programming more broadly.

There are several interesting and promising avenues for future work. Larger crowd sizes would enable more extensive evaluation of the solutions submitted on MLPro. MLPro can also be used directly in recruiting for ML-related industry and academic positions, in contrast to the current standard of asking algorithms-based coding questions in conjunction with theoretical ML questions. The existing approaches used in industry for ML interviewing notably leaves out an evaluation of on-the-job skills. Systems like MLPro can be evaluated with respect to job outcomes of participants.

**ACKNOWLEDGMENTS**

PW would like to acknowledge support from Mr. Schroeder and the Stanford Interdisciplinary Graduate Fellowship (SIGF) as the Schroeder Family Goldman Sachs Graduate Fellow. MLPro acknowledges funding from Contour Venture Partners, Cardinal Ventures, and FAST.XYZ.

# REFERENCES

[1] Aaron, Samuel, and Alan F. Blackwell. "From sonic Pi to overtone: creative musical experiences with domain-specific and functional languages." In *Proceedings of the first ACM SIGPLAN workshop on Functional art, music, modeling & design*, pp. 35-46. 2013.

[2] Anderson, David P., Jeff Cobb, Eric Korpela, Matt Lebofsky, and Dan Werthimer. "SETI@ home: an experiment in public-resource computing." *Communications of the ACM* 45, no. 11 (2002): 56-61.

[3] Balderas, Antonio, JUAN MANUEL Dodero, Manuel Palomo-Duarte, and Ivan Ruiz-Rube. "A domain specific language for online learning competence assessments." *International Journal of Engineering Education* 31, no. 3 (2015): 851-862.

[4] Beberg, Adam L., Daniel L. Ensign, Guha Jayachandran, Siraj Khaliq, and Vijay S. Pande. "Folding@ home: Lessons from eight years of volunteer distributed computing." In *2009 IEEE International Symposium on Parallel & Distributed Processing*, pp. 1-8. IEEE, 2009.

[5] Belsley, David A., Edwin Kuh, and Roy E. Welsch. Regression diagnostics: Identifying influential data and sources of collinearity. John Wiley & Sons, 2005.

[6] Carreira-Perpinán, Miguel A. "A review of dimension reduction techniques." *Department of Computer Science. University of Sheffield. Tech. Rep. CS-96-09* 9 (1997): 1-69.

[7] Chen, Daniel. "A Pedagogical Approach to Create and Assess Domain-Specific Data Science Learning Materials in the Biomedical Sciences." PhD diss., Virginia Tech, 2022.

[8] Das, Rhiju, Benjamin Keep, Peter Washington, and Ingmar H. Riedel-Kruse. "Scientific discovery games for biomedical research." *Annual Review of Biomedical Data Science* 2 (2019): 253-279.

[9] Eiben, Christopher B., Justin B. Siegel, Jacob B. Bale, Seth Cooper, Firas Khatib, Betty W. Shen, Barry L. Stoddard, Zoran Popovic, and David Baker. "Increased Diels-Alderase activity through backbone remodeling guided by Foldit players." *Nature biotechnology* 30, no. 2 (2012): 190-192.

[10] Hadi Mogavi, Reza, Xiaojuan Ma, and Pan Hui. "Characterizing Student Engagement Moods for Dropout Prediction in Question Pool Websites." Proceedings of the ACM on Human-Computer Interaction 5, no. CSCW1 (2021): 1-22.

[11] Harrison Jr, David, and Daniel L. Rubinfeld. "Hedonic housing prices and the demand for clean air." *Journal of environmental economics and management* 5, no. 1 (1978): 81-102.

[12] Isdahl, Richard, and Odd Erik Gundersen. "Out-of-the-box reproducibility: A survey of ML platforms." In *2019 15th international conference on eScience (eScience)*, pp. 86-95. IEEE, 2019.

[13] Jing, Longlong, and Yingli Tian. "Self-supervised visual feature learning with deep neural networks: A survey." *IEEE transactions on pattern analysis and machine intelligence* (2020).

[14] Johnson, Jeremiah W., and Karen H. Jin. "Jupyter notebooks in education." *Journal of Computing Sciences in Colleges* 35, no. 8 (2020): 268-269.

[15] Kafai, Yasmin B., Mike Horn, Joshua A. Danish, Megan Humburg, Xintian Tu, Bria Davis, Chris Georgen et al. "Affordances of Digital, Textile and Living Media for Designing and Learning Biology in K-12 Education." In *ICLS*. 2018.

[16] Klimeš, Jonáš. "Domain-Specific Language for Learning Programming." (2016).

[17] Lam, Amy T., Jonathan Griffin, Matthew Austin Loeun, Nate J. Cira, Seung Ah Lee, and Ingmar H. Riedel-Kruse. "Pac-Euglena: A Living Cellular Pac-Man Meets Virtual Ghosts." In *Proceedings of the 2020 CHI Conference on Human Factors in Computing Systems*, pp. 1-13. 2020.

[18] Lam, Amy T., Karina G. Samuel-Gama, Jonathan Griffin, Matthew Loeun, Lukas C. Gerber, Zahid Hossain, Nate J. Cira, Seung Ah Lee, and Ingmar H. Riedel-Kruse. "Device and programming abstractions for spatiotemporal control of active micro-particle swarms." *Lab on a Chip* 17, no. 8 (2017): 1442-1451.

[19] Kawrykow, Alexander, Gary Roumanis, Alfred Kam, Daniel Kwak, Clarence Leung, Chu Wu, Eleyine Zarour et al. "Phylo: a citizen science approach for improving multiple sequence alignment." *PloS one* 7, no. 3 (2012): e31362.

[20] Kim, Sunyoung, Christine Robson, Thomas Zimmerman, Jeffrey Pierce, and Eben M. Haber. "Creek watch: pairing usefulness and usability for successful citizen science." In *Proceedings of the SIGCHI Conference on Human Factors in Computing Systems*, pp. 2125-2134. 2011.

[21] Lee, Jeehyung, Wipapat Kladwang, Minjae Lee, Daniel Cantu, Martin Azizyan, Hanjoo Kim, Alex Limpaecher et al. "RNA design rules from a massive open laboratory." *Proceedings of the National Academy of Sciences* 111, no. 6 (2014): 2122-2127.

[22] Lintott, Chris J., Kevin Schawinski, Anže Slosar, Kate Land, Steven Bamford, Daniel Thomas, M. Jordan Raddick et al. "Galaxy Zoo: morphologies derived from visual inspection of galaxies from the Sloan Digital Sky Survey." *Monthly Notices of the Royal Astronomical Society* 389, no. 3 (2008): 1179-1189.

[23] Maguire, Phil, Rebecca Maguire, and Robert Kelly. "Using automatic machine assessment to teach computer programming." Computer Science Education 27, no. 3-4 (2017): 197-214.

[24] Pedregosa, Fabian, Gaël Varoquaux, Alexandre Gramfort, Vincent Michel, Bertrand Thirion, Olivier Grisel, Mathieu Blondel et al. "Scikit-learn: Machine learning in Python." *the Journal of machine Learning research* 12 (2011): 2825-2830.

[25] Salihbegovic, Adnan, Teo Eterovic, Enio Kaljic, and Samir Ribic. "Design of a domain specific language and IDE for Internet of things applications." In *2015 38th international convention on information and communication technology, electronics and microelectronics (MIPRO)*, pp. 996-1001. IEEE, 2015.

[26] Silaparasetty, Nikita. "An Overview of ML." ML Concepts with Python and the Jupyter Notebook Environment (2020): 21-39.

[27] Simpson, Robert, Kevin R. Page, and David De Roure. "Zooniverse: observing the world's largest citizen science platform." In *Proceedings of the 23rd international conference on world wide web*, pp. 1049-1054. 2014.

[28] Sorzano, Carlos Oscar Sánchez, Javier Vargas, and A. Pascual Montano. "A survey of dimensionality reduction techniques." *arXiv preprint arXiv:1403.2877* (2014).

[29] Washington, Peter, Cezmi Onur Mutlu, Aaron Kline, Kelley Paskov, Nate Tyler Stockham, Brianna Chrisman, Nick Deveau, Mourya Surhabi, Nick Haber, and Dennis P. Wall. "Challenges and Opportunities for ML Classification of Behavior and Mental State from Images." *arXiv preprint arXiv:2201.11197* (2022).

[30] Washington, Peter, Karina Samuel-Gama, and Ingmar Riedel-Kruse. "A Programming Toolkit for Automating Biophysics Experiments with Microorganism Swarms." *Biophysical Journal* 114, no. 3 (2018): 183a.

[31] Washington, Peter, Karina G. Samuel-Gama, Shirish Goyal, and Ingmar H. Riedel-Kruse. "Bioty: A cloud-based development toolkit for programming experiments and interactive applications with living cells." *bioRxiv* (2017): 236919.

[32] Washington, Peter, Karina G. Samuel-Gama, Shirish Goyal, Ashwin Ramaswami, and Ingmar H. Riedel-Kruse. "Interactive programming paradigm for real-time experimentation with remote living matter." *Proceedings of the National Academy of Sciences* 116, no. 12 (2019): 5411-5419.

[33] Washington, Peter, Karina G. Samuel-Gama, Shirish Goyal, Ashwin Ramaswami, and Ingmar H. Riedel-Kruse. "Prototyping biotic games and interactive experiments with javaScript." In *Extended Abstracts of the 2018 CHI Conference on Human Factors in Computing Systems*, pp. 1-4. 2018.

[34] Wasik, Szymon, Maciej Antczak, Jan Badura, Artur Laskowski, and Tomasz Sternal. "A survey on online judge systems and their applications." ACM Computing

Surveys (CSUR) 51, no. 1 (2018): 1-34.

[35] Zahari, Ana Syafiqah, Lukman Ab Rahim, Nur Aisyah Nurhadi, and Mubeen Aslam. "A Domain-Specific Modelling Language for Adventure Educational Games and Flow Theory." *International Journal onn Advanced Science Engineering Information Technology* 10, no. 3 (2020): 999-1007.

[36] Zinovieva, I. S., V. O. Artemchuk, Anna V. Iatsyshyn, O. O. Popov, V. O. Kovach, Andrii V. Iatsyshyn, Y. O. Romanenko, and O. V. Radchenko. "The use of online coding platforms as additional distance tools in programming education." In Journal of Physics: Conference Series, vol. 1840, no. 1, p. 012029. IOP Publishing, 2021.
8